\documentclass{JHEP3}
\usepackage{amsmath}
\usepackage{epsfig}
\usepackage{cite}

\newcommand{\as}{\alpha_s}
\newcommand{\asb}{\bar{\alpha}_s}
\newcommand{\CF}{C_F}

\newcommand{\di}{\mathrm{d}}
\newcommand{\mc}[1]{\mathcal{#1}}
\newcommand{\mr}[1]{\mathrm{#1}}

\newcommand{\akt}{\mathrm{anti-k_t}}
\newcommand{\kt}{k_t}
\newcommand{\ca}{\mathrm{C/A}}

\title{On the resummation of clustering logarithms for non--global observables}

\author{Yazid Delenda \\
        D\'{e}partement des Sciences de la Mati\`{e}re, Facult\'{e} des
        Sciences\\ Universit\'{e} Hadj Lakhdar - Batna, Algeria\\
        \email{yazid.delenda@yahoo.com}}

\author{Kamel Khelifa-Kerfa \\
        School of Physics and Astronomy, University of Manchester,\\
        Oxford road, Manchester M13 9PL, U.K.\\
        \email{Kamel.Khelifa@hep.manchester.ac.uk}}

\preprint{MAN/HEP/2012/10}

\abstract{Clustering logs have been the subject of much study in
recent literature. They are a class of large logs which arise for
non-global jet-shape observables where final-state particles are
clustered by a non-cone--like jet algorithm. Their resummation to
all orders is highly non--trivial due to the non-trivial role of
clustering amongst soft gluons which results in the phase-space
being non-factorisable. This may therefore significantly impact the
accuracy of analytical estimations of many of such observables.
Nonetheless, in this paper we address this very issue for jet shapes
defined using the $k_t$ and C/A algorithms, taking the jet mass as our
explicit example. We calculate the coefficients of the Abelian
$\alpha_s^2 L^2$, $\alpha_s^3 L^3$ and $\alpha_s^4 L^4$ NLL terms in
the exponent of the resummed distribution and show that the impact
of these logs is small which gives confidence on the perturbative
estimate without the neglected higher-order terms. Furthermore we
numerically resum the non-global logs of the jet mass distribution
in the $k_t$ algorithm in the large-$N_c$ limit.}


\begin{document}

\section{Introduction}
\label{sec:Intro}

Event and jet shapes have played a significant role in the
development and testing of QCD and its parameters (see ref.
\cite{Dasgupta:2003iq} for a review), ranging from the confirmation
of vector-boson nature of gluons \cite{QCD} to very accurate
measurements of the coupling and colour factors of its underlying
theory \cite{Bethke:2002rv,Beneke:1998ui}, as well as tuning Monte
Carlos \cite{Corcella:2000bw,Sjostrand:2006za,MC2,substructure1}.
They have also been used to extract moments of the non-perturbative
coupling, as reviewed in refs. \cite{Dasgupta:2003iq} and \cite{Beneke:1998ui} (and references therein).

Despite the fact that the LEP collider was a clean environment to
``measure'' QCD using event/jet shapes, they will still be an
important tool at the LHC, particularly jet shapes and jet
substructure which are expected to be invaluable tools in the search
for new physics due to the final state being dominantly jets
\cite{MC2,substructure1}. For instance there has recently been much
interest in the concept of substructure of jets, used to identify
massive boosted electroweak objects whose hadronic decay products
tend to cluster into the same fat jet, e.g, 
\cite{substructure1,substructure2,substructure3,substructure4,
substructure5,substructure6}. The invariant jet mass, being central
to jet shapes and phenomenologically most useful, provides a simple
way of signal/background discrimination. Whilst the jet-mass
distribution of QCD jets is, away from the Sudakov peak (at low values), featureless, that of heavy-particles
decay products has ``bumps'' \cite{substructure4}.

Given their unequivocal importance at the LHC, event/jet-shape
observables have received substantial progress over the last few
years, both at fixed order and to all orders in the perturbative
expansion. For ``global'' observables, Next-to--Leading Log (NLL)
resummation, NLO fixed-order calculations and their matching have
been available for a number of event/jet shapes for quite a while
\cite{Catani:1992ua, Banfi:2001bz}. In fact, calculations of
arbitrary such observables can now be performed automatically using
programs such as \texttt{CAESAR} \cite{CEASAR}. The
state-of--the-art perturbative calculations is up to NNLO for fixed
order \cite{NNLO} and up to N$^3$LL accuracy\footnote{We speak of
the logarithmic accuracy in the exponent of the distribution.} for
resummation \cite{Catani:1992ua, Banfi:2001bz}. Moreover,
non-perturbative calculations have seen noticeable advancements,
especially in disentangling various components of these effects such
as the underlying event and hadronisation \cite{Dassalmag}.

However there is a class of observables termed ``non-global''
\cite{ngl1,ngl2} which suffer from large logs, which are absent for global observables and which have not been \emph{fully} resummed even at
NLL. The invariant jet mass, with or without a cut on inter-jet
energy flow, is a typical example. Resummation of non-global logs,
valid only in the large-$N_c$ approximation, has long been available
for a wide range of observables which are linear in soft momenta and
progress has been made for jet-defined quantities which are not
linear in soft emissions, e.g.  gaps-between--jets energy flow
\cite{Appleby:2002ke, Delenda:2006nf} and jet mass with a jet veto
\cite{KhelifaKerfa:2011zu}. Furthermore it was found in
\cite{Appleby:2002ke, Delenda:2006nf} and \cite{KhelifaKerfa:2011zu}
that applying a clustering algorithm on the final-state partons
seems to significantly restrict the phase-space responsible for
non-global logs, thus producing a sizeable reduction in their
phenomenological impact.

In the presence of a jet algorithm, other than anti-$k_t$
\cite{antikt}, non-global observables suffer from large logs in the
Abelian part of the emission amplitude.
These are referred to as ``clustering logs''
\cite{KhelifaKerfa:2011zu,Kelley:2012kj,Kelley:2012zs}. The
clustering logs were first computed at fixed order in ref. \cite{BD05} and
subsequently \emph{partially} resummed in ref. \cite{Delenda:2006nf}
for away-from--jets energy flow. There, however, the resulting
exponent of the resummed distribution has been written as a
power-series in the radius parameter of the jet algorithm ($R$)
starting from $R^3$. Thus for a typical jet radius ($R\sim 1$) it
was sufficient, for an accurate approximation, to compute the first
couple of terms ($\mathcal{O}(R^3) $ and $\mathcal{O}(R^5)$), as the
series rapidly converge. Excellent agreement was noticed when
compared to the output of the Monte Carlo (MC) developed in
\cite{ngl1}.

Clustering logs arise due to mis-cancellation between real emissions
and virtual corrections. This mis-cancellation results from re-clustering of
final-state configurations of soft gluons. Unlike the single-log
$E_t$ distribution, resummation of clustering logs in the jet mass
distribution cannot simply be written as a power-series in the jet
radius. Collinear singularities at the boundary of a small-$R$ jet
yields large logs in the radius parameter, which appear to all
orders in $\alpha_s$ \cite{KhelifaKerfa:2011zu,BDKM}. Note that the
jet veto distribution, studied in the latter references,
disentangles from the jet mass distribution to all orders
\cite{0303101} and has a non-global structure analogous to the $E_t$
distribution. That is, the coefficients of both non-global and
clustering logs are identical for the jet veto and $E_t$
distributions. The arguments of the logs are different though
\cite{KhelifaKerfa:2011zu}.

The clustering logs have recently been the subject of much study
both at fixed order and to all orders. However there has been no
\emph{full} resummation to all orders and it was recently suggested
that it is unlikely that such logs be fully resummed even to their
leading log level, which means NLL accuracy relative to leading
double logs \cite{Kelley:2012kj,Kelley:2012zs}. In this paper we
address this very issue for the jet mass distribution.

In ref. \cite{BDKM} the resummation of the jet mass distribution for $e^+e^-$ annihilation, in
the anti-$k_t$ algorithm \cite{antikt}, was performed to all orders
including the non-global component in large-$N_c$ limit. Employing
the anti-$k_t$ clustering algorithm meant that the observable
definition was linear in transverse momenta of soft emissions and
therefore the resummation involved no clustering logs. In the same
paper the authors also carried out a fixed-order calculation for the
jet mass distribution at $\mathcal{O}(\alpha_s^2)$ employing the
$k_t$ algorithm \cite{inclukt,ESkt} in the small-$R$ limit. The
result of integration yielded an NLL term, $\alpha_s^2 L^2$, that is
formally as important as Sudakov primary NLL terms.

In the framework of soft-collinear effective theory \cite{SEFT} the
authors of ref. \cite{Kelley:2012kj} confirmed the findings of ref.
\cite{BDKM} and computed the full $R$-dependence of the jet mass
distribution at $\mathcal{O}(\alpha_s^2)$. They also pointed out the
unlikelihood of the resummation of the  clustering logs to all
orders.

We present here an expression for the all-orders resummed result of
clustering logs to NLL accuracy in the $k_t$ and C/A algorithms
\cite{inclukt,Caachen} for the jet mass distribution. The logs that
we control take the form $\mc{F}_n\alpha_s^n L^n$ ($n\geq 2$) in the
exponent of the resummed distribution, and we only compute
$\mc{F}_2$, $\mc{F}_3$ and $\mc{F}_4$ in this paper. By comparing
our findings to the output of the Monte Carlo of ref. \cite{ngl1} we
show that missing higher-order terms are negligible. We also show
that the impact of the primary-emission single clustering logs is of
maximum order 5\% for typical jet radii. Furthermore we estimate the
non-global (clustering-induced) contribution to the jet mass
distribution in the large-$N_c$ limit using the Monte Carlo of ref.
\cite{ngl1} in the case of the $k_t$ algorithm.

This paper is organised as follows. In the next section we define
the jet mass and show how different algorithms affect its
distribution at NLL level. We then start with the impact of $k_t$
and C/A algorithms on the jet mass distribution at $\mc O
(\alpha_s^2)$, thus confirming the findings of ref. \cite{BDKM}. In
section \ref{sec:3-4loop} we unearth higher order clustering terms
and notice that they exhibit a pattern of exponentiation. By making
an anstaz of higher-order clustering coefficients we perform a
resummation of the clustering logs to all orders in sec.
\ref{sec.all-orders} and compare our findings to the output of a
Monte Carlo program in sec. \ref{sec:MC}. We also perform a
numerical estimate of the non-global logs in the large-$N_c$ limit
for the jet mass distribution in the $k_t$ algorithm in sec.
\ref{Sec:NG}. Finally we draw our conclusions and point to future
work.

\section{The jet mass distribution and clustering algorithms}
\label{sec.jm distrib}

Consider for simplicity the $e^+e^-$ annihilation into two jets
produced back-to-back with high transverse momenta. We would like to
study the single inclusive jet mass distribution via measuring the
invariant mass of one of the final-state jets, $M_j^2$, while
leaving the other jet unmeasured. One can restrict inter-jet
activity by imposing a cut $Q_0$ on emissions in this region
\cite{Ellis:2009wj, Ellis:2010rwa}. For the purpose of this paper we
do not worry about this issue because the effect of this cut has
been dealt with in the literature \cite{ Delenda:2006nf,
KhelifaKerfa:2011zu, BDKM} and can be included straightforwardly.

The normalised invariant jet-mass--squared fraction, $\rho$,  is
defined by:
\begin{equation}
\rho = \left(\sum_{i\in j} p_i\right)^2/ \left(\sum_i
E_{i}\right)^2, \label{eq:RhoDefinition}
\end{equation}
where the sum in the numerator runs over all particles in the
measured jet, defined using an infrared and collinear (IRC)-safe
algorithm such as $k_t$ or C/A algorithm \cite{inclukt,Caachen}. At
born level the jet mass has the value zero and it departs from this
value at higher orders.

In general sequential recombination algorithms
\cite{antikt,inclukt,Caachen,SRA1} one defines distance measures for
each pair of objects\footnote{Objects refer to actual tracks, cells
and towers in the detector and to particles/partons in perturbative
calculations.} $\{i,j\}$ in the final state as \cite{SRA1}:
\begin{equation}
d_{ij} =  2\min\left(E_i^{2p}, E_j^{2p}\right)
\left(1-\cos\theta_{ij}\right), \label{eq:dijMeasure}
\end{equation}
and for each parton $i$ a distance, from the beam $B$,
\begin{equation}
d_{iB} = E_i^{2p}\,R^2. \label{eq:diMeasure}
\end{equation}
The values $p=-1,0,1$ correspond respectively to the anti-$k_t$, C/A
and $k_t$ algorithms. A typical algorithm takes the smallest value
of these distances and merges particles $i$ and $j$ into a single
object when $d_{ij}$ is the smallest, using an appropriate
combination scheme such as addition of four-momenta. If instead a
distance $d_{iB}$ is the smallest then object $i$ is considered a
jet and is removed from the list of final-state objects. This
procedure is then iterated until all objects have been clustered
into jets. We may write $2(1-\cos\theta_{ij})\approx \theta_{ij}^2$
in the small-angles limit where jets are narrow and well-separated
to avoid correlations, and hence contamination, between various
jets.

In the $k_t$ algorithm, softest partons are clustered first
according to the above mentioned procedure, while in the anti-$k_t$
algorithm clustering starts with the hardest partons. In the C/A
algorithm only angular separations between partons matters so
clustering starts with the geometrically closest partons. In effect,
clustering induces modifications to the mass of the measured jet due
to reshuffling of soft gluons. Only those gluons which end up in the
jet region would contribute to its mass. Different jet algorithms
would then give different values of the jet mass for the same event.

The global part of the integrated resummed jet mass distribution in
the $k_t$ (C/A) algorithm is related to that in the anti-$k_t$
algorithm by:
\begin{equation}
\Sigma^{k_t (\ca)}\left(\frac{R^2}{\rho}\right) =
\Sigma^{\akt}\left(\frac{R^2}{\rho}\right)
\exp\left[{g_{2,A}^{k_t(\ca)}\left(\frac{R^2}{\rho}\right)}\right],
\label{eq:GeneralResummedFormkt-CA}
\end{equation}
where $\Sigma^{\akt}$ resums the leading double logs (DL) (due to
soft and collinear poles of the emission amplitude) as well as
next-to--leading single logs (SL) in the anti-$k_t$ algorithm. The
reader is referred to ref. \cite{BDKM} for further details about
this piece. The function $g_{2,A}$ contains the new large clustering
single logs due to $k_t$(C/A) clustering.

In addition to the global part, each of the distributions
$\Sigma^{\akt, k_t, \ca}$ receives its own non-global NLL contribution
factor $\mc S(R^2/\rho)$. In the anti-$k_t$ algorithm the
resummation of the non-global logs in the large-$N_c$ limit for the jet mass
distribution was estimated in ref. \cite{BDKM}. The result was actually shown to coincide with that of the hemisphere jet mass case. The latter has been available for quite a while \cite{ngl1, ngl2}. For $k_t$ clustering
the effect of non-global logs has been dealt with in the literature,
for example, in the case of gaps-between--jets $E_t$ flow
\cite{Appleby:2002ke, Delenda:2006nf}. We expect the gross features
of the latter to hold for the jet mass observable. Essentially, the
impact of $k_t$ clustering is in such a way as to reduce the size of
the anti-$k_t$ (non-clustering) non-global logs. Brief comments on
such a reduction will be given in sec. \ref{Sec:NG}. In this paper
we are, however, mainly interested in the function
$g_{2,A}(R^2/\rho)$.

\section{Two-gluon emission calculation}
\label{sec:2-gluon} The effect of $k_t$ and C/A clustering logs
starts at $\mathcal{O}(\alpha_s^2)$. At $\mathcal{O}(\alpha_s)$
there is indeed a dependence on the jet radius but this dependence
is also present for the anti-$k_t$ algorithm and has been dealt with
in ref. \cite{BDKM}. In this regard we begin with two-gluon emission
case and study the effect of clustering using both the C/A and $k_t$
clustering algorithms. This calculation has already been performed
in ref. \cite{BDKM} for the $k_t$ algorithm in the small-$R$ limit.
Here we perform a full-$R$ calculation of the logs coefficient.

\subsection{Calculation in the $k_t$ algorithm}

Consider the independent emission of two soft energy-ordered gluons
$k_2\ll k_1\ll Q$, where $Q$ is the hard scale. This regime, i.e.
strong energy ordering, is sufficient to extract the leading
clustering logs.

In the case of the anti-$k_t$ algorithm discussed in ref.
\cite{BDKM} the only contribution from these gluons to the jet mass
differential distribution is when both of them are emitted within an
angle $R$ from the hard parton initiating the measured jet. This is
because the anti-$k_t$ algorithm works in the opposite sense of the
$k_t$ algorithm: clustering starts with the hardest particles. In
this sense the algorithm essentially works as a perfect cone around
the hard initiating parton with no dragging-in or dragging-out
effect.

In the $k_t$ algorithm, on the other hand, the two gluons may be in
one of the following four configurations:
\begin{enumerate}
\item Both gluons $k_1$ and $k_2$ are initially, i.e. before applying
the algorithm, inside\footnote{We use inside, or simply ``in'', to
signify that the parton is within an angular separation of $R$ from
the hard triggered parton (jet axis), and outside, or simply
``out'', if it is more than $R$ away from it.} the triggered jet.
This configuration contributes to the jet shape (mass) regardless of
clustering. The corresponding contribution to the jet mass
distribution is identical to that of the anti-$k_t$ case (accounted
for by $\Sigma^\akt$ -- see eq. \eqref{eq:C2generalFormula} below).
\item Both gluons are initially outside the triggered jet. This arrangement
does not contribute to the jet shape regardless of whether the two
gluons are clustered or not.
\item The harder gluon, $k_1$, is initially inside the triggered jet
and the softer gluon, $k_2$, is outside of it. The value of the jet
shape (mass) is not changed even if clustering takes place. This is
due to the strong-ordering condition stated above.
\item \label{Config} The harder gluon, $k_1$, is initially
outside the triggered jet and the softer gluon, $k_2$, is inside of
it. Applying the algorithm one finds that a real-virtual
mis-cancellation occurs only  if $k_2$ is pulled-out of the
triggered-jet vicinity by $k_1$, a situation which is only possible
if $k_2$ is ``closer'' (in terms of distances $d_{ij}$) to $k_1$
than to the axis of the jet.
\end{enumerate}
We translate the latter configuration mathematically into the step
function:
\begin{equation}
\Xi_{2}\left(k_1, k_2\right) = \Theta(\theta_1^2-R^2)
\Theta(R^2-\theta_2^2) \Theta(\theta_2^2-\theta_{12}^2),
\label{eq:ClustFun2}
\end{equation}
where $\theta_i$ is the angle between gluon $k_i$ and the jet axis
and $\theta_{12}$ is the relative angle of the two gluons. The above
condition is valid only in the small-$R$ limit and we extend this to
the full $R$-dependence in appendix \ref{Sec:F2}. Hence $\Xi_2 = 1$
when configuration \ref{Config} above is satisfied and $\Xi_2 = 0$
otherwise. While in the case where the gluon $k_1$ is virtual and
$k_2$ is real (meaning that $k_1$ cannot pull $k_2$ out) the
particle $k_2$ contributes to the jet mass. To the contrary, when
both $k_1$ and $k_2$ are real, then $k_1$ will not allow $k_2$ to
contribute to the jet mass as it pulls it out. Thus a real-virtual
mismatch occurs and a tower of large logarithms appears. To
calculate these large logarithms we insert the clustering condition
$\Xi_2$ above into the phase-space of the $\mc{O}(\alpha_s^2)$
integrated jet mass distribution. The latter, normalised to the Born
cross-section $\sigma_0$ is given, in the soft and collinear
approximation, by:
\begin{eqnarray}
\nonumber \Sigma_2^{\kt}\left(R^2/\rho\right)
&=& \Sigma_2^\akt\left(R^2/\rho\right) + \Sigma_2^{\mathrm{clus}}\left(R^2/\rho\right),\\
\Sigma_2^{\mathrm{clus}}\left(R^2/\rho\right) &=&
\frac{1}{2!}\left(-C_F \frac{\alpha_s}{\pi}\right)^2 \int  dP_1 dP_2
\,\Xi_2\left(k_1,k_2\right), \label{eq:C2generalFormula}
\end{eqnarray}
with
\begin{equation}
dP_i = \frac{d\omega_i}{\omega_i}
\frac{d\cos\theta_i}{\sin^2\theta_i} \frac{d\phi_i}{\pi}
\Theta\left(\frac{4\omega_i}{Q} (1-\cos\theta_i)-\rho\right) \approx
\frac{dx_i}{x_i} \frac{d\phi_i}{2\pi} \frac{d\theta_i^2}{\theta_i^2}
\Theta(x_i \theta_i^2-\rho), \label{eq:DiffPhaseSpace2}
\end{equation}
where $\theta_i$, $\phi_i$ and $x_i = 2\,\omega_i/Q$ are the polar
angles, with respect to the jet axis, and the energy fraction of the
$i^{\mathrm{th}}$ gluon. The factor $1/2!$ in eq.
\eqref{eq:DiffPhaseSpace2} compensates for the fact that we are
considering both orderings: $x_2 \ll x_1 \ll Q$ and $x_1 \ll x_2 \ll
Q$. Had we chosen to work with only one ordering, say $x_2 \ll x_1
\ll Q$ as stated at the beginning of this section, then the
said-factor would have not been included. In terms of the
coordinates $(\theta, \phi, x)$ the jet mass fraction defined in eq.
\eqref{eq:RhoDefinition} reduces to (in the small-angles limit):
\begin{equation}
\rho = \frac{4\omega_i}{Q} (1-\cos\theta_i) \approx x_i\,\theta_i^2,
\label{eq:rhoDefined}
\end{equation}
where gluon $i$ is in the jet. In eq. \eqref{eq:DiffPhaseSpace2} we
used the step function to restrict the jet mass instead of
Dirac-$\delta$ function because we are considering the integrated
distribution instead of the differential one. Hence at two-gluon
level, the correction term due to $k_t$ clustering (eq.
\eqref{eq:C2generalFormula}) is given by:
\begin{multline}
\Sigma_2^{\mathrm{clus}} = \frac{1}{2!}
\left(-\frac{C_F\alpha_s}{\pi}\right)^2 \int^1 \frac{dx_1}{x_1}
\frac{dx_2}{x_2} \int_0
\frac{d\theta_1^2}{\theta_1^2}\frac{d\theta_2^2}{\theta_2^2}
\int_{-\pi}^{\pi} \frac{d\phi_1}{2\pi} \Theta(x_1\theta_1^2-\rho)
\times \\
\times \Theta(x_2\theta_2^2-\rho) \Theta(\theta_1^2-R^2)
\Theta(R^2-\theta_2^2) \Theta(\theta_2^2-\theta_{12}^2),
\label{eq:C2Distr}
\end{multline}
where we used our freedom to set $\phi_2$ to 0. We write the result
to single-log accuracy as:
\begin{equation}
\Sigma_2^{\mathrm{clus}} =
\frac{1}{2!}\left(-\frac{C_F\alpha_s}{\pi}\right)^2 \mathcal{F}_2\,
L^2, \label{eq:C2}
\end{equation}
where $L \equiv \ln\left(R^2/\rho\right)$. In the above equation we
ignored subleading logs and used the fact that in the small-angles
approximation: $\theta_{12}^2 = \theta_1^2
+\theta_2^2-2\theta_{1}\theta_2\cos\phi_1$. The two-gluon
coefficient $\mc F_2$ is given by:
\begin{equation}
\mathcal{F}_2 =  \frac{2}{\pi} \int _0^{\frac{\pi }{3}}  d\phi \ln^2
(2 \cos \phi) =\frac{\pi^2}{54} \approx 0.183. \label{eq:F2Coeff}
\end{equation}

As stated at the outset of this section, we can actually compute
$\mc F_2$ beyond the small-angles (thus small-$R$) limit. In
appendix \ref{Sec:F2} we present an analytic calculation of this
clustering coefficient as an expansion in the radius parameter. One
observes that the small-$R$ approximation of $\mc F_2$, given in eq.
\eqref{eq:F2Coeff}, is actually valid for jet radii up to order
unity because of its slow variation with $R$ (first correction to
the small-$R$ result is of $\mc O (R^4)$). For instance, at $R =
0.7$ and $R = 1.0$ the coefficient is $\mc F_2 = 0.188$ and $0.208$
respectively, i.e. an increment of about $3\%$ and $15\%$. We
compare the analytical formula with the full numerical result in
Fig. \ref{fig:F2Full}.

\subsection{Calculation in the C/A algorithm}

At the two-gluon level, order $\as^2$ in the perturbative expansion
of the shape distribution, the C/A and $k_t$ algorithms work
essentially in a similar manner. Although the C/A algorithm clusters
partons according only to the polar distances between the various
pairs (recall $p=0$ in Eqs. \eqref{eq:dijMeasure} and
\eqref{eq:diMeasure}), at this particular level energies do not seem
to play a role (once they are assumed strongly-ordered). The jet
shape is only altered if the softer gluon $k_2$ is (geometrically)
closer to $k_1$, or vice versa for the opposite ordering, than to
the jet axis, so as to escape clustering with the jet. This
similarity between the two algorithms does not hold to all orders
though. We shall explicitly show, in the next section, that they
start differing at $\mc O(\as^3)$.

\section{Three and four-gluon emission} \label{sec:3-4loop}

\subsection{Three-gluon emission}

\subsubsection{$k_t$ clustering case}

Consider the emission of three energy-ordered soft gluons $Q \gg
k_1\gg k_2 \gg k_3$. We proceed in the same way as for the
two-gluons case. First we write the step function $\Xi_3(k_1,
k_2,k_3)$, which describes the region of phase-space that gives rise
to clustering logs. To this end, applying the $k_t$ clustering
algorithm yields the following expression for $\Xi_3$:
\begin{eqnarray}
\nonumber \Xi_3 (k_1,k_2,k_3)  &=&
\Theta(R^2-\theta_3^2)\Theta(\theta_2^2-R^2)\Theta(R^2-\theta_1^2)
\Theta(\theta_3^2-\theta_{23}^2)+ k_1 \leftrightarrow k_2+\\
\nonumber &&
+\Theta(R^2-\theta_3^2)\Theta(\theta_2^2-R^2)\Theta(\theta_1^2-R^2)
\Theta(\theta_3^2-\theta_{23}^2)\Theta(\theta_3^2-\theta_{13}^2)+\\
&&
+\Theta(R^2-\theta_3^2)\Theta(R^2-\theta_2^2)\Theta(\theta_1^2-R^2)
\Theta(\theta_{13}^2-\theta_3^2)
\Theta(\theta_{23}^2-\theta_3^2)\Theta(\theta_{2}^2-\theta_{12}^2).\nonumber\\
\label{eq:ClustFun3kt}
\end{eqnarray}
Hence the correction term due to $k_t$ clustering at three-gluon
level, $\Sigma^{\mathrm{clus}}_3$, which is of an analogous form to
$\Sigma^{\mathrm{clus}}_2$ (eq. \eqref{eq:C2generalFormula}), is:
\begin{multline}
\Sigma^{\mathrm{clus}}_3 = \frac{1}{3!}
\left(-\frac{C_F\alpha_s}{\pi}\right)^3 \int^1 \frac{dx_1}{x_1}
\frac{dx_2}{x_2} \frac{dx_3}{x_3} \int_0
\frac{d\theta_1^2}{\theta_1^2} \frac{d\theta_2^2}{\theta_2^2}
\frac{d\theta_3^2}{\theta_3^2} \int_{-\pi}^{\pi}
\frac{d\phi_1}{2\pi} \frac{d\phi_2}{2\pi} \times \\ \times
\Theta(x_1\theta_1^2-\rho)\Theta(x_2\theta_2^2-\rho)\Theta(x_3\theta_3^2-\rho)
\times \Xi_3(\theta_1,\theta_2,\theta_3,\phi_1,\phi_2),
\label{eq:C3Distrkt}
\end{multline}
where, as in $\Sigma^{\mathrm{clus}}_2$, we used our freedom to set
$\phi_3=0$. Performing the integration, in the small-$R$ limit, the
final result may be cast, to single-log accuracy, in the form:
\begin{equation}
\Sigma^{\mathrm{clus}}_3 =
\frac{1}{3!}\left(-\frac{C_F\alpha_s}{\pi}\right)^3 \left[
\frac{3\times 2}{2} \frac{L^2}{2} \mathcal{F}_2 L^2 + \mathcal{F}_3
L^3\right], \label{eq:C3kt}
\end{equation}
where the three-gluon coefficient $\mathcal{F}_3 = -0.052$.
Extending the formalism developed in appendix \ref{Sec:F2} to the
case of three gluons, it is possible to write down an expansion of
$\mc F_3$ in terms of $R$, just as we did with $\mc F_2$. However,
since (a) the whole clustering logs correction to the anti-$k_t$
result is substantially small, as we shall see later in sec.
\ref{sec:MC}, and (b) $|\mc F_3|$ is much smaller than $\mc F_2$, we do
not perform such an analytical calculation here. We do perform a numerical
evaluation of the full--$R$ dependence of $\mc F_3$ for various values of $R$, though. The final
results are provided in table \ref{tab:ClusCoeffNumericalValues}.

The first leading term in eq. \eqref{eq:C3kt} is the product of the
one-gluon DL leading term in the anti-$k_t$ algorithm, $\alpha_s
L^2$, and the two-gluon SL term of eq. \eqref{eq:C2}. The second NLL
term in eq. \eqref{eq:C3kt} is the new clustering log at $\mc
O(\alpha_s^3)$. We expect that at $n^{\mathrm{th}}$ order in
$\alpha_s$ new clustering logs of the form $\mc F_n L^n$ emerge.
Notice that $|\mc F_3| < \mc F_2$, indicating that the series $\mc
F_n L^n$ rapidly converges. Consequently, the two-gluon result
is expected to be the dominant contribution. This expectation will
be strengthened in the next section, where we compute the four-gluon
coefficient $\mc F_4$. Before doing so, we address the three-gluon
calculation in the C/A algorithm.

\subsubsection{C/A algorithm case}

We follow the same procedure, outlined above for the $k_t$
algorithm, to extract the large logarithmic corrections to the
anti-$k_t$ result in the C/A algorithm. As we stated before the
algorithm deals with the angular separations of partons only. In
this regard we consider the various possibilities of the angular
configurations of gluons and apply the algorithm accordingly, to
find a mis-cancellation of real-virtual energy-ordered soft
emissions. The clustering condition step function in the C/A
algorithm can be expressed as:
\begin{equation}
\Xi_3^\ca = \Xi_3^{\kt} +  \widetilde{\Xi}_3, \label{eq:ClustFun3CA}
\end{equation}
where $\Xi_3^{\kt}$ is given in eq. \eqref{eq:ClustFun3kt} and the
extra function reads:
\begin{equation}
\widetilde{\Xi}_3 = \Theta(R^2-\theta_3^2) \Theta(\theta_1^2-R^2)
\Theta(\theta_{13}^2-\theta_3^2)\Theta(\theta_3^2-\theta_{23}^2)
\Theta(\theta_2^2-\theta_{12}^2)
\Theta(\theta_{23}^2-\theta_{12}^2). \label{eq:ClustFun3CAExtra}
\end{equation}
Inserting the step function  eq. \eqref{eq:ClustFun3CA} into the
equivalent of eq. \eqref{eq:C3Distrkt} for the C/A algorithm one
obtains:
\begin{equation}
\Sigma^{\mathrm{clus}, \ca}_3 = \Sigma^{\mathrm{clus}}_3 +
\frac{1}{3!} \left(-\frac{C_F\alpha_s}{\pi}\right)^3
\widetilde{\mathcal{F}}_3\,L^3, \label{eq:C3ca}
\end{equation}
where $\Sigma_3^\mathrm{clus}$ is given in eq. \eqref{eq:C3kt} and $ \widetilde{\mathcal{F}}_3 = 0.0236$. Hence the factor
$\mathcal{F}_3^{\mathrm{C/A}}$ which replaces the $k_t$ clustering
term $\mathcal{F}_3$ is, in the small-$R$ limit,
$\mathcal{F}_3^{\mathrm{C/A}} = \mathcal{F}_3 +
\widetilde{\mathcal{F}}_3 = -0.028$. We provide the full-$R$
numerical estimates of $\mc F^\ca_3$ in table
\ref{tab:ClusCoeffNumericalValues}. This result illustrates that the
contribution to the shape distribution at this order ($\as^3$) in
the C/A clustering is approximately half that in the $k_t$
clustering  (although the values are, in both algorithms,
substantially small). Moreover, it confirms the conclusion
reached-at in the $k_t$ algorithm case, for the C/A algorithm,
namely that the clustering logs series are largely dominated by the
two-gluon result. Next, we present the calculation of the four-gluon
coefficient $\mc F_4$.

\subsection{Four-gluon emission}

Consider  the emission of four energy-ordered soft primary gluons $Q
\gg k_1 \gg k_2 \gg k_3 \gg k_4$. The determination of the
clustering function $\Xi_4$ is more complex than previous lower
orders, particularly for the C/A algorithm. As a result of this we
only present, in this paper, the findings for the $k_t$ algorithm.
The four-gluon calculations for the jet mass variable are very much
analogous to those presented in ref. \cite{Delenda:2006nf} for the
energy flow distribution, to which the reader is referred for
further details. Here we confine ourselves to reporting on the final
answers. After performing the necessary phase-space integration,
which is again partially carried out using Monte Carlo integration
methods, and simplifying one arrives at the following expression for
the correction term, to the anti-$k_t$ shape distribution, due to
$k_t$ clustering:
\begin{equation}
\Sigma^{\mathrm{clus}}_4 = \frac{1}{4!}
\left(-\frac{C_F\alpha_s}{\pi}\right)^4\times\bigg \{
    6  \times \frac{L^4}{4}  \times  \mathcal{F}_2 L^2
+     4  \times    \frac{L^2}{2}\times \mathcal{F}_3 L^3+ 3 \times
(\mathcal{F}_2 L^2)^2 + \mathcal{F}_4 L^4 \bigg\}, \label{eq:C4kt}
\end{equation}
where, in the small-$R$ limit, $\mathcal{F}_4 = 0.0226$. The
full-$R$ numerical results are presented in table
\ref{tab:ClusCoeffNumericalValues}. As anticipated earlier we have
$\mc F_4 < |\mc F_3| \ll \mc F_2$, thus confirming the rapid
convergence behaviour of the clustering logs series.

eq. \eqref{eq:C4kt} contains products of terms in the expansion of
the Sudakov anti-$k_t$ form factor with the two- and three-gluon
results, Eqs. \eqref{eq:C2} and \eqref{eq:C3kt}, as well as the new
NLL clustering term at $\mc O(\as^4)$. The first leading term in eq.
\eqref{eq:C4kt} comes from the product $\as^2 L^4\times$ two-gluon
result ($\as^2 \mc F_2 L^2$); the second term comes from the product
$\as L^2\times$ three-gluon ($\as^3 \mc F_3 L^3$) result; and the
third term is the square of the two-gluon result. Therefore, the
three- and four-gluon expressions, Eqs. \eqref{eq:C3kt} and
\eqref{eq:C4kt}, seem to suggest a pattern of ``exponentiation''.
Such behaviours give rise to the intriguing possibility of finding a reasonably
good approximation to the full resummation of clustering logs to all
orders, the task to which we now turn.

\section{All-orders result}\label{sec.all-orders}

In analogy to the work of ref. \cite{Delenda:2006nf}, one can see
that the results obtained at 2, 3 and 4-gluon can readily be
generalised to $n$-gluon level, with new terms of the form
$\mathcal{F}_n^{(\ca)} L^n$ appearing at each order for the $k_t$
(C/A) algorithm. By similar arguments to those of ref.
\cite{Delenda:2006nf} we can deduce the leading term in the
$n^{\mathrm{th}}$ order contribution due the $k_t$ (C/A) clustering
to the shape distribution, $\Sigma_n^{\mathrm{clus}(\ca)}$. It
reads:
\begin{equation}
\Sigma_n^{\mathrm{clus}(\ca)} \propto\,
\frac{1}{(n-2)!}\,\left(-\frac{C_F\alpha_s }{\pi}
\frac{L^2}{2}\right)^{n-2} \frac{\mathcal{F}_2}{2}
\left(-\frac{C_F\alpha_s }{\pi} L\right)^{2}   ,\qquad n\geq 2
\label{eq:CnF2Leading}.
\end{equation}
Summing up the terms $\Sigma_n^{\mathrm{clus}(\ca)}$ to all orders,
i.e. from $n=2$ to $n \rightarrow \infty$, yields the following
resummed expression:
\begin{equation}
\Sigma^{\mathrm{clus}(\ca)} \propto \, \exp\left\{
-\frac{C_F\alpha_s}{\pi} \frac{ L^2}{2} \right\}
\frac{\mathcal{F}_2}{2} \left(-\frac{C_F \alpha_s}{\pi}\,
L\right)^2. \label{eq:SigmaClusA}
\end{equation}
The first exponential in the above expression is the celebrated
Sudakov form factor, that one obtains when resumming the jet mass
distribution in the anti-$k_t$ algorithm. Due to its Abelian nature,
the Sudakov is entirely determined by the first primary emission
result. There are also other pure $\mc F_2$ terms in
$\Sigma_n^{\mathrm{clus}(\ca)}$ for $ n \geq 4$ of the form:
\begin{equation}
\Sigma_n^{\mathrm{clus}(\ca)} \propto\,
\frac{1}{(n-4)!}\left(-\frac{C_F\alpha_s}{\pi}\frac{L^2}{2}
\right)^{n-4}\frac{\mathcal{F}_2^2}{8}
\left(-\frac{C_F\alpha_s}{\pi} L \right)^4 ,
\label{eq:CnF2SubLeading}
\end{equation}
which can be resummed to all orders into:
\begin{equation}
\Sigma^{\mathrm{clus}(\ca)} \propto\, \exp\left\{ -\frac{C_F\alpha_s
}{\pi} \frac{L^2}{2} \right\} \frac{\mathcal{F}_2^2}{8}
\left(-\frac{C_F\alpha_s}{\pi} L \right)^4. \label{eq:SigmaClusB}
\end{equation}
From Eqs. \eqref{eq:SigmaClusA} and \eqref{eq:SigmaClusB}, one
anticipates the resummed result to all orders in the clustering log,
$L$, to be of the form:
\begin{equation}
\Sigma^{\mathrm{clus}(\ca)} \propto  \exp\left\{ -\frac{C_F\alpha_s
}{\pi} \frac{L^2}{2} \right\} \left[ \exp\left\{\frac{1}{2}\,\mc
F_2\, \left(-\frac{C_F \as}{\pi}\, L\right)^2\right\}  -1 \right].
\label{eq:SigmaClusC}
\end{equation}

Furthermore we have the following expression in
$\Sigma_n^{\mathrm{clus}(\ca)}$:
\begin{equation}
\Sigma_n^{\mathrm{clus}(\ca)} \propto\,
\frac{1}{(n-3)!}\,\left(-\frac{C_F\alpha_s }{\pi}
\frac{L^2}{2}\right)^{n-3} \frac{1}{6} \mathcal{F}_3^{(\ca)}
\left(-\frac{C_F\alpha_s }{\pi} L\right)^{3}   ,\qquad n\geq 3
\label{eq:CnF3Leading}.
\end{equation}
which is resummed into:
\begin{equation}
\Sigma^{\mathrm{clus}(\ca)} \propto \, \exp\left\{
-\frac{C_F\alpha_s}{\pi} \frac{ L^2}{2} \right\}
\frac{\mathcal{F}_3^{(\ca)}}{6} \left(-\frac{C_F \alpha_s}{\pi}\,
L\right)^3. \label{eq:F3res}
\end{equation}

Similarly, one expects analogous expressions to eq.
\eqref{eq:SigmaClusC} for the remaining $\mc F_3$, $\mc F_4$,
$\cdots$ terms, in addition to ``interference terms'' between these
coefficients, e.g. $\mc F_2 \mc F_3 $ which should first show up at
$\mc{O} (\alpha_s^5L^5)$. Recall that the shape distribution in the
$k_t$ (and C/A) algorithm is, in the Abelian primary emission part,
the sum of the distribution in the anti-$k_t$ algorithm
(clustering-free distribution) and a clustering-induced
distribution. Schematically:
\begin{equation}
\Sigma^{\kt(\ca)} = \Sigma^{\akt} + \Sigma^{\rm{clus}(\rm{clus},\ca)},
\end{equation}
where $\Sigma^\akt$ is simply the Sudakov form factor mentioned
above. Thus gathering everything together and including the logs
which are present in the anti-$k_t$ case the following
exponentiation is deduced:
\begin{equation}
\Sigma^{\kt(\ca)} =  \exp\left\{ -\frac{C_F\alpha_s}{\pi} \frac{
L^2}{2} \right\} \exp\left\{\sum_{n\geq 2} \frac{1}{n!}\,
\mathcal{F}_n^{(\ca)}\left(-\frac{C_F\alpha_s}{\pi} L \right)^n
\right\}, \label{eq:SigmaAbeliankt-CA}
\end{equation}
where $\mc F_n^{(\ca)}$ is the $n^{\mathrm{th}}$-gluon coefficient
in the $\kt$ (C/A) algorithm.

Confined to the anti-$k_t$ jet algorithm, the authors in \cite{BDKM}
computed the full resummed jet mass distribution up to NLL accuracy,
including the effect of the running coupling as well as hard
collinear emissions\footnote{Leaving non-global contributions aside
for now.}. Taking these and the fixed-order loop-constants into
account, eq. \eqref{eq:SigmaAbeliankt-CA} becomes:
\begin{equation}
\Sigma^{\kt(\ca)} = \left(1+\sum_n c_n \asb^n\right)\exp\left[ L g_1
(\alpha_s L) + g_2(\alpha_s L) \right] \exp
\left[g_{2,A}^{\kt(\ca)}(\alpha_s L)\right],
\label{eq:SigmaAbelianFullkt}
\end{equation}
where $\asb = \as/2\pi$ and the functions $g_1$ and $g_2$ resum the leading and
next-to--leading logs occurring in the anti-$k_t$ case. Their
explicit formulae are given in \cite{BDKM}. The new piece in the
resummation which is due to primary-emission clustering and which
contributes at NLL level is:
\begin{equation}
g_{2,A}^{\kt(\ca)}(\as\,L)=\sum_n \frac{1}{n!}\,
\mathcal{F}_n^{(\ca)}\left(-2\,\CF\,t\right)^n,
\label{eq:g2AFinalForm}
\end{equation}
where we have introduced the evolution parameter $t$, which governs
the effect of the running coupling:
\begin{equation}
t = \frac{1}{2\pi} \int_{Q\sqrt{\rho}/R}^{Q} \frac{\di k_t}{k_t}\,
\as(k_t) = -\frac{1}{4\pi \beta_0} \ln(1-\as\beta_0 L),
\end{equation}
where the last equality is the one-loop expansion of $t$, and we
have $\beta_0 = (11C_A - 2n_f )/12\pi$.

In the present work, we have been able to compute, by means of brute
force, the first three coefficients, $\mc F_{i=2,3,4}$, for the
$k_t$ algorithm and only the first two coefficients, $\mc
F_{i=2,3}^{\ca}$, for the C/A algorithm. They are, nonetheless,
sufficient to capture the behaviour of the all-orders result, for a
range of jet radii, as we shall show in the next section where we
compare our findings to the output of a numerical Monte Carlo
program.

\section{Comparison to MC results} \label{sec:MC}

The MC program we use was first developed in \cite{ngl1} to resum
non-global logs in the large-$N_c$ limit, and later modified to
include the $k_t$ clustering in \cite{Appleby:2002ke, Delenda:2006nf}. Since the MC
program was originally designed to resum soft wide-angle emissions
to all orders, it only resums single logs. The leading logs in the
jet mass distribution are, however, double logs. As such one cannot
produce the corresponding DL Sudakov form factor with the MC. Hence
it is not possible to directly compare the output of the MC with eq.
\eqref{eq:SigmaAbelianFullkt}, in order to verify the analytical
calculations of $g_{2,A}$. One can, however, extract the MC resummed
clustering function, $\exp\left[g_{2,A}^{\mr{MC}}\right]$, by
subtracting off the result with clustering ``switched off'' from
that with clustering ``switched on''. The remainder is then directly
compared to $\exp\left[g_{2,A}\right]$, where $g_{2,A}$ is given in
eq. \eqref{eq:g2AFinalForm}. Such comparisons are presented in fig.
\ref{fig:CLs_1jet_R1-R0.1}. 
\FIGURE[ht]{
\epsfig{file=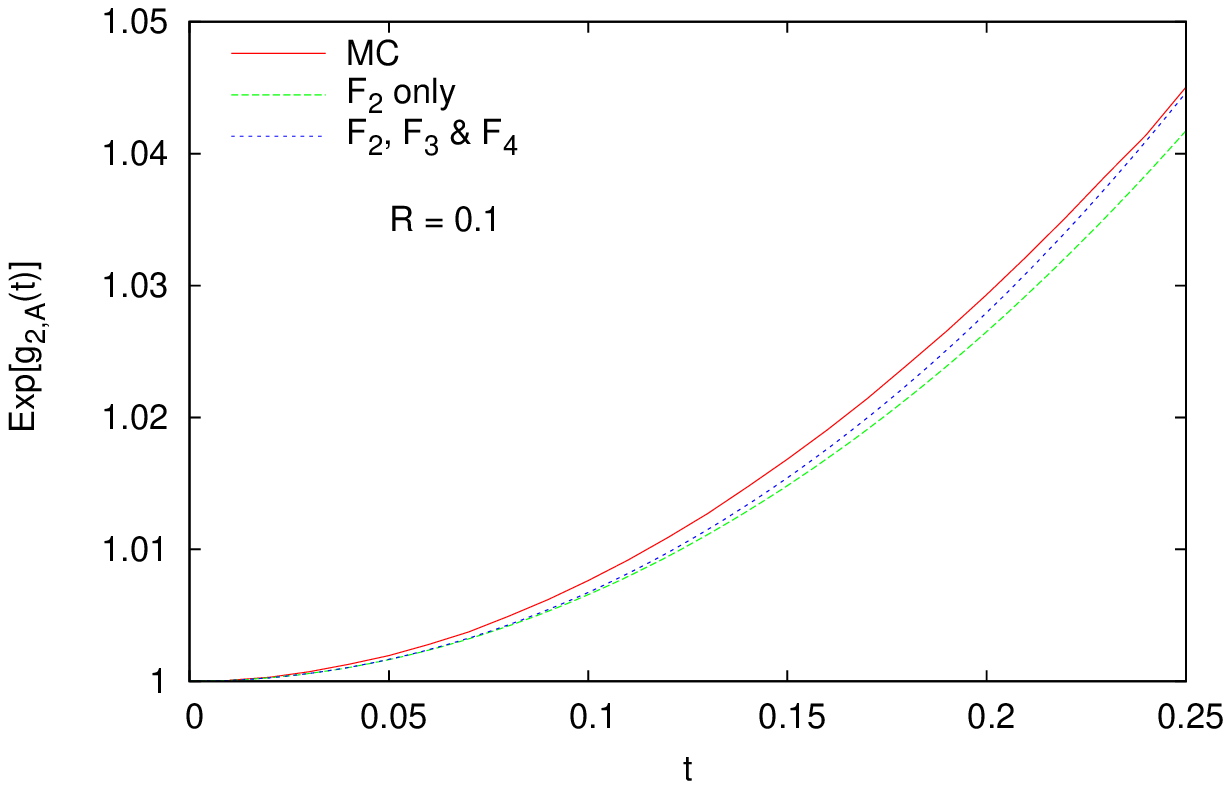,width=0.7\textwidth}
\epsfig{file=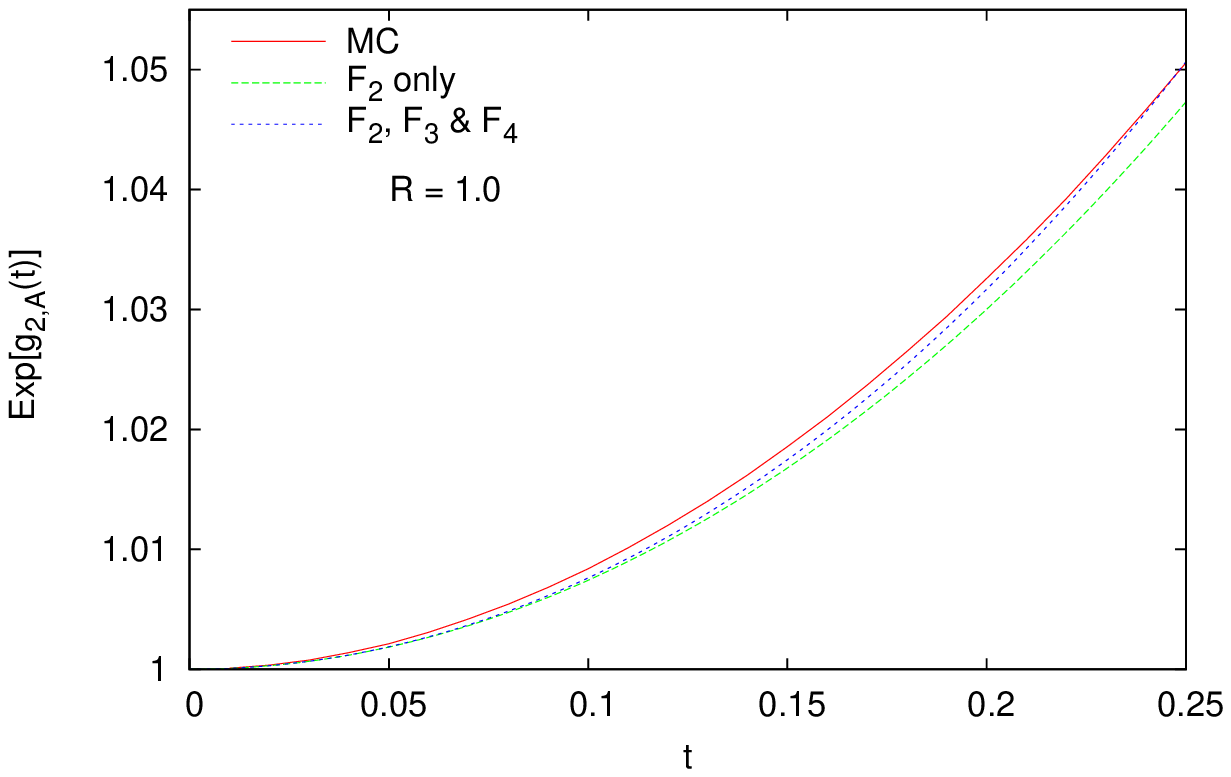,width=0.7\textwidth} 
\caption{Comparisons
of the analytical result to the output of the Monte Carlo program in
the $k_t$ algorithm for two values of the jet
radius.\label{fig:CLs_1jet_R1-R0.1}}}

The plots display the MC estimate of $\exp\left(g_{2,A}\right)$, the
analytical  result of the latter in the cases where (a) only the
first coefficient $\mc F_2$ is included in the sum
\eqref{eq:g2AFinalForm} and (b) the first three coefficients, $\mc
F_{i=2,3,4}$, are included. The dependence of the clustering
coefficients $\mc F_i$ on $R$, given in table \ref{tab:ClusCoeffNumericalValues}, is taken into consideration.

One can clearly see that the function $\exp\left(g_{2,A}\right)$ is
largely dominated by the first coefficient $\mc F_2$, with minor
corrections from $\mc F_3$ and $\mc F_4$. For instance for $R=1.0$
the $\mc F_2$, $\mc F_3$ and $\mc F_4$ coefficients (put alone in
turn) induce a correction  to the anti-$k_t$ resummed distribution
of 1.6\%, 0.06\%, 0.002\% and 4.8\%, 0.3\% and 0.02\% for $t=0.15$ and $0.25$ respectively. This can be
understood from eq. \eqref{eq:g2AFinalForm}: in addition to being
smaller than $\mc F_2$, the higher-gluon coefficients $\mc F_n$
($n\geq 3$) are suppressed by a factorial factor ($n!$), thus
leading to a fast convergence. Given the agreement between our
analytical estimate and the output of the Monte Carlo, and given
that the function $\exp(g_{2,A})$ contributes at most $\mc O(5\%)$,
we conclude that our results for the resummed clustering logs are
phenomenologically accurate for jet radii up to order unity, and
that missing higher-order coefficients $\mc F_n$ ($n\geq 5$) are
unimportant.

Lastly, we plot in fig. \ref{fig:CLs_1jet_g2_AllR}, the MC results
of the function $\exp (g_{2,A})$ for various jet radii. The plots
unequivocally indicate that for a fixed $t$, say $0.15$, the function
$g_{2,A}$ varies very slowly with $R$ for $R\leq 1$ and grows
relatively rapidly as $R> 1$. Such a behaviour may be explained by the
analytical formula of $\mc F_2$, eq. \eqref{eq:F2RResult}, where the first
correction to the small--$R$ result is proportional to $R^4$. 

\FIGURE[ht]
{\epsfig{file=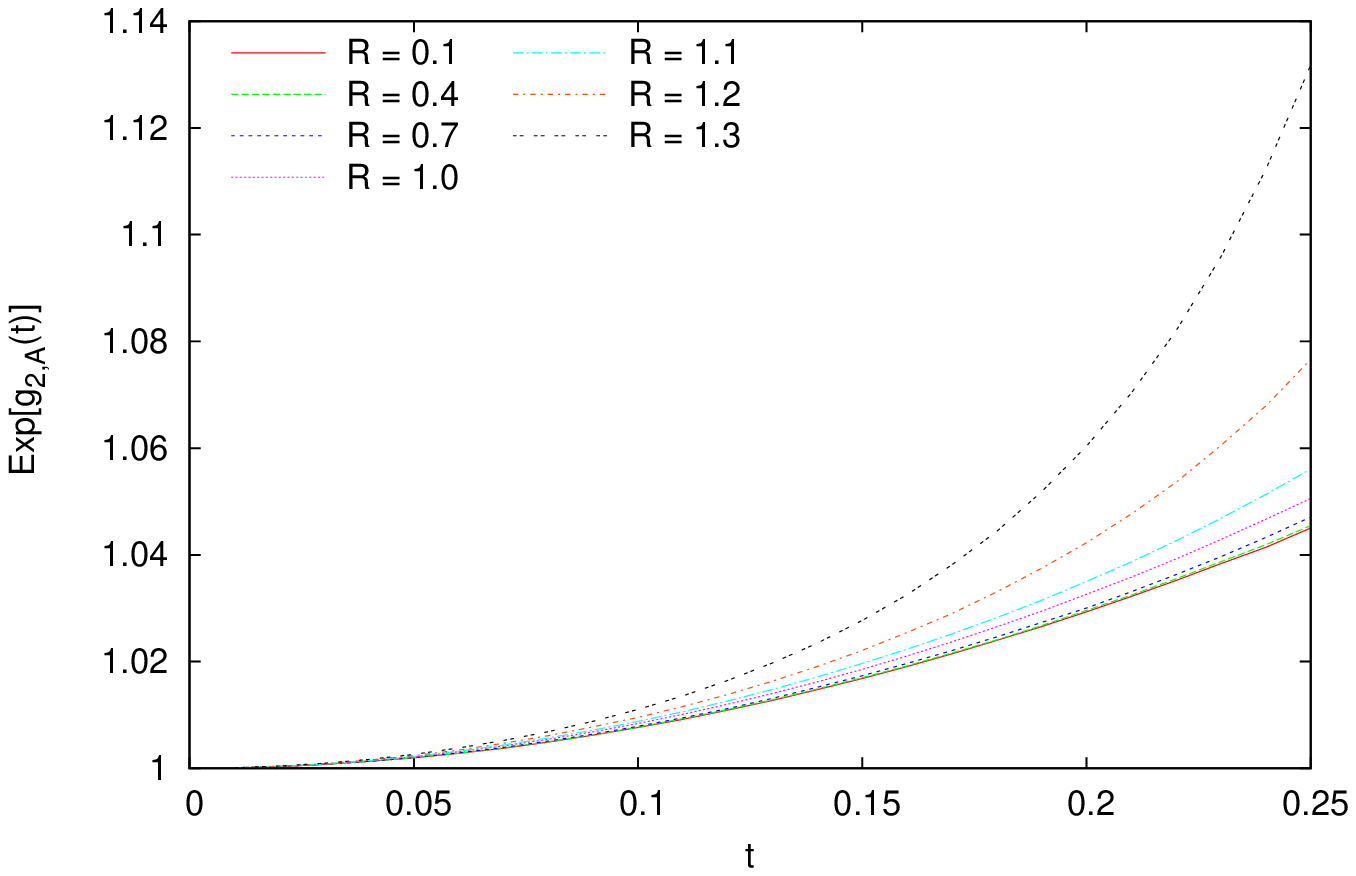,width=0.7\textwidth}
\caption{The output of the Monte Carlo program in the $k_t$
algorithm for various jet radii.
\label{fig:CLs_1jet_g2_AllR}}}

\subsection{Non-global logs}

\label{Sec:NG}

\FIGURE{
\epsfig{file=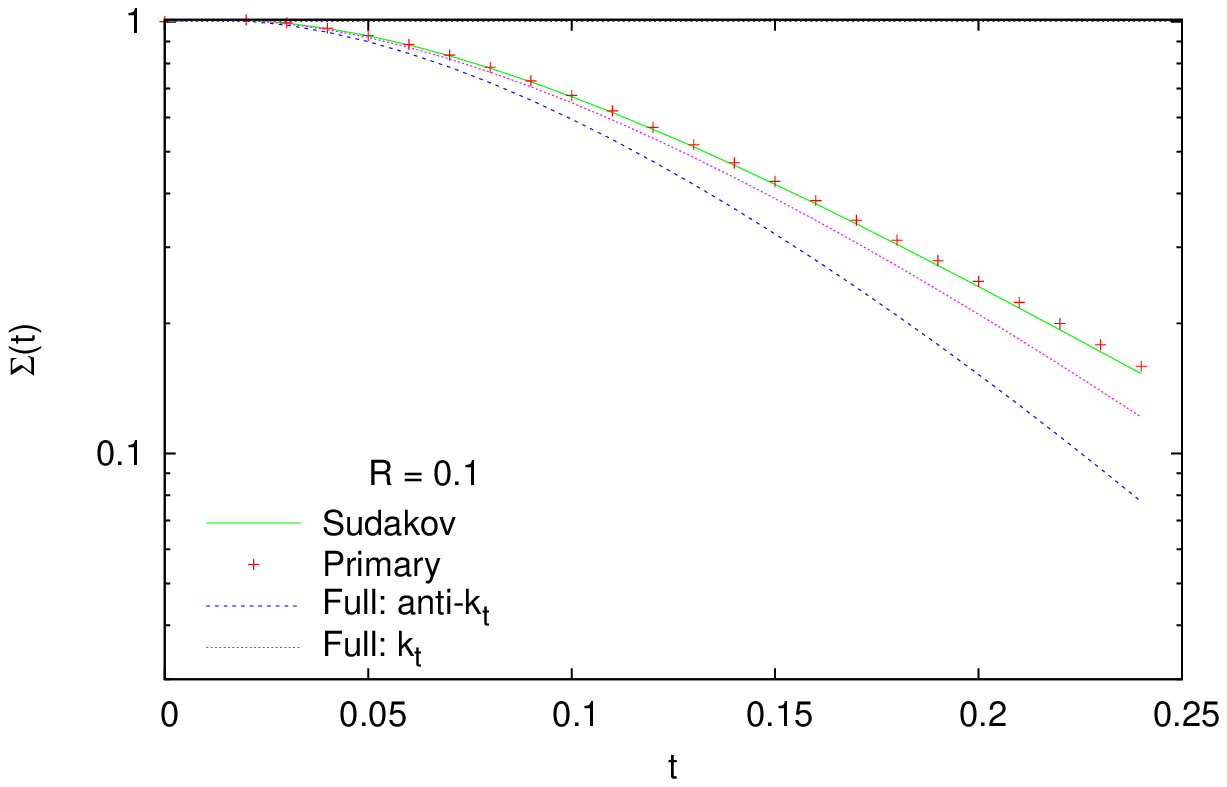,width=0.7\textwidth}
\epsfig{file=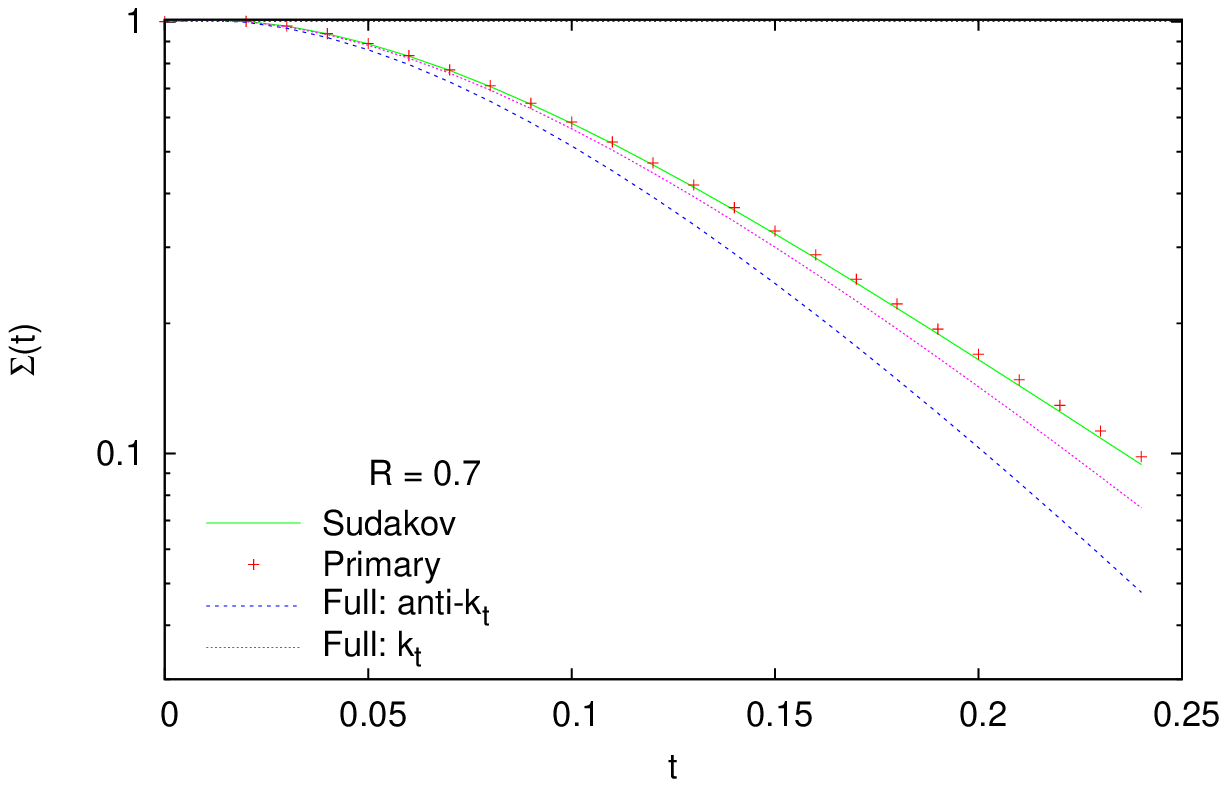,width=0.7\textwidth}
\epsfig{file=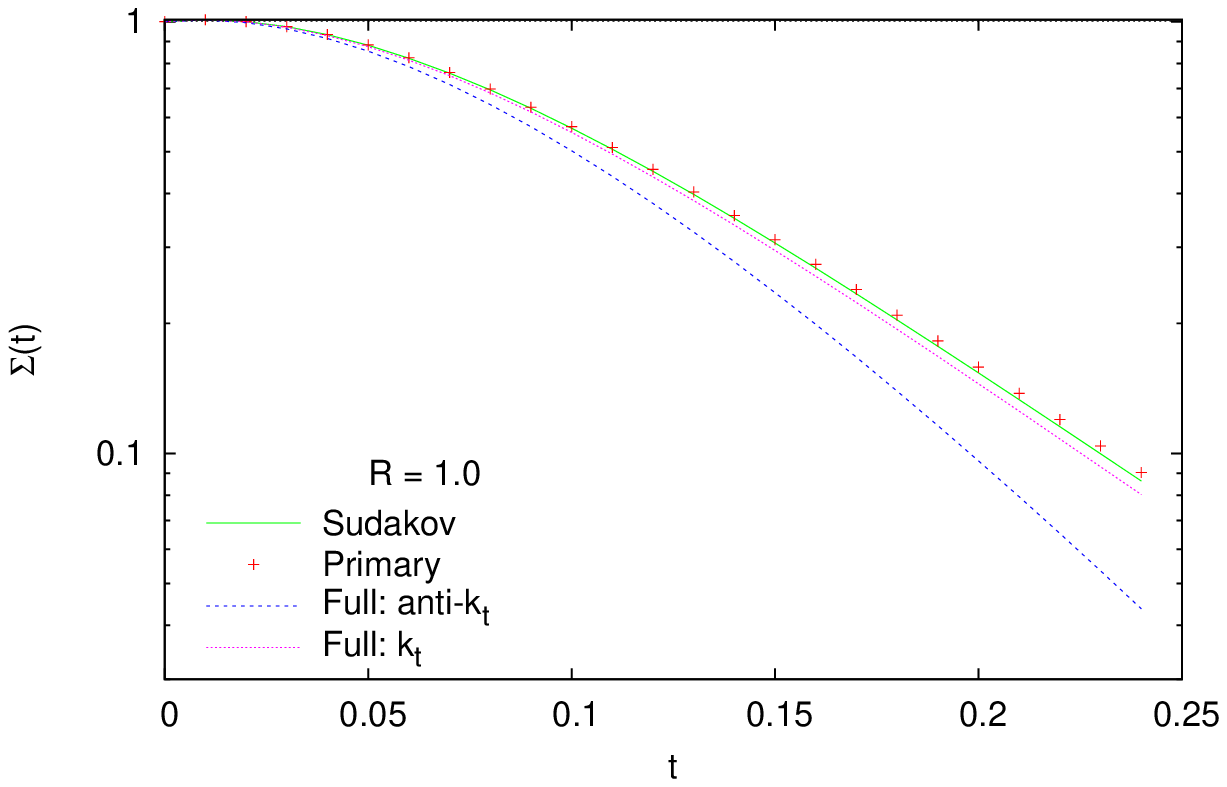,width=0.7\textwidth}
\caption{\label{fig:FullResumJetMassDistktB}Comparisons of the
Sudakov result, the correct primary result and the full result
including non-global logarithms with and without clustering, as
detailed in the main text.} }

As a final task, we plot in Fig. \ref{fig:FullResumJetMassDistktB}
the full resummed jet mass distribution in the $k_t$ algorithm, eq.
\eqref{eq:GeneralResummedFormkt-CA}, including the non-global factor
$\mc S(t)$ in the large-$N_c$ limit, for various jet radii. In the
figure, ``Sudakov'' refers to the Sudakov anti-$\kt$ form factor,
$\Sigma^\akt$, ``primary'' refers to the primary form factor in the
$\kt$ algorithm containing the clustering logarithms,
$\Sigma^{\kt}$ (eq. \eqref{eq:SigmaAbelianFullkt}), ``Full: anti-$\kt$'' refers to the full anti-$\kt$
resummation, $\Sigma^\akt \, \mc S^\akt(t)$ and ``Full: $\kt$''
refers to $\Sigma^{\kt} \, \mc S^{\kt}(t)$. We notice that the
inclusion of non-global logs leads to a noticeably large reduction
of the full resummed result for the anti-$\kt$ algorithm case, while
in the $k_t$ algorithm the reduction is moderate.

As is well known by now \cite{Appleby:2002ke, Delenda:2006nf,
KhelifaKerfa:2011zu, Hornig:2011tg}, clustering reduces the impact
of non-global logs through restricting the available phase-space for
their contribution. This is also the case with the jet mass
distribution where we note, for instance, that for $t = 0.15$ the
impact of non-global logs in the anti-$k_t$ algorithm (for $R=0.1, 0.7$
and $1.0$) is a 24\% reduction of the global part, while in the$k_t$
clustering case this is merely 7\% (for $R=0.1$ and $R=0.7$) and 4\%
for $R=1.0$. We further notice that for small values of the jet radius, $\mc S(t)$ is independent of $R$.
While this is true for all values of $R$ in the anti--$\kt$ algorithm \footnote{$\mc S^\akt(t)$ for our jet mass is identical to that for the hemisphere jet mass case considered in \cite{ngl1}. }, in the $\kt$ algorithm
$\mc S$ falls down as $R$ becomes larger. This is evident in the $R=1.0$ plot in Fig. \ref{fig:FullResumJetMassDistktB}.

\section{Conclusions}\label{sec.conc}

In this paper we have considered the possibility of exponentiation
of clustering logs in the $k_t$ and C/A algorithms. We have found,
by explicit calculations of the first few orders (up to $\mc
O(\as^4)$, and including, for the first time in literature, the full
jet radius dependence), that the perturbative expansion of the
invariant jet mass distribution exhibits a pattern of an expansion
of an exponential. Consequently we were able to write an all-orders
\emph{partially} resummed expression for primary emission clustering
logs. We further checked our formula against the output of a
numerical Monte Carlo and found a good agreement, within the
accuracy of our calculations. We have therefore concluded that
missing higher-order single-log terms in our resummed distribution
have a negligible impact on the total resummed distribution for
typical values of jet radii (up to order unity).

Furthermore, we have briefly discussed the impact of the inclusion
of non-global logs on the total resummed distribution. We confirmed
previous observations concerning the facts that (a) non-global logs
reduce the Sudakov peak of the distribution and that (b) such an
impact is diminished when clustering is imposed on final-state
particles.

We note that the calculations we performed here can readily be
generalised to a large class of non-global observables  defined
using the $k_t$ or C/A algorithm, where the observable is sensitive
to soft emissions in a restricted region of phase space, e.g. angularities \cite{Ellis:2010rwa}. Since the calculations of the coefficients $\mc F_n$
presented here are in fact independent of the jet shape and depend
only on the angular configurations introduced by the clustering
algorithm, then the effect of jet clustering can simply be included
for any generic observable $v$ (being sensitive to soft and
collinear emissions inside the jet only) by introducing the
exponential function $\exp[g_{2,A}]$, with exactly the same
coefficients $\mc F_n$ we computed here. The only difference is the
argument of the logarithm essentially becoming $R^2/v$.

Although we have confined ourselves to studying the single jet mass
distribution in $e^+e^-$ annihilation, the extension of our work to,
e.g. monojet production at the LHC (such as $Z$+ jet, which is an
important channel in looking at BSM physics, and for which a
calculation in the anti-$k_t$ algorithm has recently been performed
\cite{Z1jet}) can readily be performed. In \cite{prep} we consider
the extension of the work of ref. \cite{Z1jet}, which is carried out
in the anti-$k_t$ jet algorithm, to the case where final-state jets
are defined in the $k_t$ (and C/A) jet algorithms. We employ the
techniques developed for $e^+e^-$ colliders in this paper to compute
the full $R$-dependent resummed clustering logs as well as
non-global logs for hadron colliders. Although the calculations for
the latter are much more involved, no major deviations from the
overall picture drawn at the current paper are anticipated.

\acknowledgments We would like to thank Mrinal Dasgupta for suggesting the current
work as well as for helpful comments on the manuscript.

\appendix

\section{Full $R$-dependence of clustering coefficients}

\label{Sec:F2}

Here we present a calculation of the dependence of the coefficient
of the clustering logs $\mc F_2$ away from the small-angles (thus
small-$R$) approximation. To do so, it is easier to work with
transverse momentum, (pseudo-)rapidity\footnote{Recall that $\eta =
-\ln \tan (\theta/2)$ and $k_t = \omega\sin\theta$.} and azimuthal
angle with respect to the \emph{beam} axis, $(k_t, \eta, \phi)$,
variables instead of energy and polar angles, as performed in sec.
\ref{sec:2-gluon}. We also specialise to the threshold limit in
which the trigged jet is created at $90^\circ$ to the beam (which is
along the $z$-axis). Our calculations can straightforwardly be
extended to the case where the triggered jet is at an arbitrary
rapidity\footnote{We perform such a calculation in
\cite{prep} for each dipole.}. We parametrise the outgoing
four-momenta as:
\begin{eqnarray}
p_1 & = & \frac{Q}{2}(1,1,0,0),\nonumber\\
p_2 & = & \frac{Q}{2}(1,-1,0,0),\nonumber\\
k_i & = & k_{ti}(\cosh\eta_i,\cos\phi_i,\sin\phi_i,\sinh\eta_i),
\end{eqnarray}
with $i = 1, 2$ for the two gluons respectively. In terms of the new
variables, the clustering function (eq. \eqref{eq:ClustFun2}) reads:
\begin{equation}
\Xi_2 (k_1,k_2) = \Theta(d_{1j}-R^2)\Theta(R^2-d_{2j})
\Theta(d_{2j}-d_{12}),
\end{equation}
with
\begin{equation}
d_{1j} = \eta_1^2+\phi_1^2,\qquad d_{2j} = \eta_2^2+\phi_2^2,\qquad
d_{12} = (\eta_1-\eta_2)^2+(\phi_1-\phi_2)^2.
\end{equation}

In this coordinate system the jet mass becomes:
\begin{equation}
\rho = \frac{2 p_1.k_i}{Q^2/4} =
\frac{4k_{ti}}{Q}(\cosh\eta_i-\cos\phi_i) = 2x_i \left(1-\frac{ \cos
\phi_i}{\cosh \eta_i}\right),
\end{equation}
when particle $k_i$ is recombined with the triggered $p_1$ jet. In
the above we expressed the jet mass in terms of the energy fraction
$x_i = 2\omega_i/Q$, with $\omega_i = k_{ti}\cosh \eta_i$.

The probability of a single virtual soft gluon correction is given
by:
\begin{equation}
d\Gamma_i = - \frac{d^3 \vec{k}_i}{2\omega_i(2\pi)^3} g_s^2 C_F
\frac{2(p_1.p_2)}{(p_1.k_i)(p_2.k_i)} = -\frac{C_F\alpha_s}{\pi}
\frac{dx_i}{x_i}d\eta\frac{d\phi_i}{\pi}
\frac{1}{\cosh^2\eta_i-\cos^2\phi_i},
\end{equation}
where we note here that the collinear limit to the triggered jet
$p_1$ corresponds to $\eta_i\to 0$ \emph{and} $\phi_i \to 0$.

Thus we can write the correction term due to clustering (eq.
\eqref{eq:C2generalFormula}) as:
\begin{equation}
\Sigma^{\rm{clus}}_2 = \frac{1}{2!}\int \prod_i^2 d\Gamma_i
\Theta\left(x_i-
\frac{\cosh\eta_i}{2(\cosh\eta_i-\cos\phi_i)}\rho\right)
\Xi_2(k_1,k_2).
\end{equation}
Performing the energy-fraction integration yields an expression
identical to eq. \eqref{eq:C2} with the full $R$-dependent
coefficient, $\mc F_2(R)$, given by:
\begin{equation}\label{eq:F2RInt}
\mc F_2(R) = \frac{1}{\pi^2}\int d\eta_1d\phi_1d\eta_2d\phi_2
\frac{1}{\cosh^2\eta_1-\cos^2\phi_1}
\frac{1}{\cosh^2\eta_2-\cos^2\phi_2} \Xi_2(k_1,k_2).
\end{equation}

This integration can be performed numerically to extract the value
of $\mc F_2$ for arbitrary $R$. However it proves useful, in the
simple case of $\mc F_2$ in order to obtain an analytic expression
at least as a power-series, to introduce the polar variables $r$ and
$\alpha$ defined by:
\begin{equation}
\eta = r \cos\alpha, \qquad \phi = r \sin\alpha.
\end{equation}
We rewrite eq. \eqref{eq:F2RInt} as:
\begin{multline}
\mc F_2 = \frac{1}{\pi^2}\int_R^\infty r_1dr_1
\int_{-\pi}^{\pi}d\alpha_1 \int_{0}^R r_2 dr_2 \int_{-\pi}^{\pi}
d\alpha_2 \Theta(2r_2\cos(\alpha_2-\alpha_1)-r_1)\times\\
\times \frac{1}{\cosh^2(r_1\cos\alpha_1)-\cos^2(r_1\sin\alpha_1)}
\,\,\frac{1}{\cosh^2(r_2\cos\alpha_2)-\cos^2(r_2\sin\alpha_2)}.
\label{eq:F2RInt2}
\end{multline}

We note that one can expand the second line of eq.
\eqref{eq:F2RInt2} in powers of $r_i$ so as to write the result as a
power-series in $R$. To first order, the second line expands as
$1/r_1^2 1/r_2^2$. We thus immediately identify this integral as the
one for the small-angles approximation, whose result is $\pi^2/54
\approx 0.183$. Performing higher-order integrals of the expansion
of the integrand yields the following result:
\begin{equation} \label{eq:F2RResult}
\mc F_2(R) = 0.183 + 0.0246 R^4 + 0.00183 R^8 + 0.000135 R^{12} +
\mc O(R^{16}).
\end{equation}
This expansion is actually valid for values of $R$ up to order
unity, a claim which is backed-up by fully performing the integral
\eqref{eq:F2RInt2} numerically via Monte Carlo methods and comparing
to the analytical estimate \eqref{eq:F2RResult}, fig.
\ref{fig:F2Full}. We note that the clustering coefficient varies
very slowly with $R$.

\FIGURE[ht]{
\epsfig{file=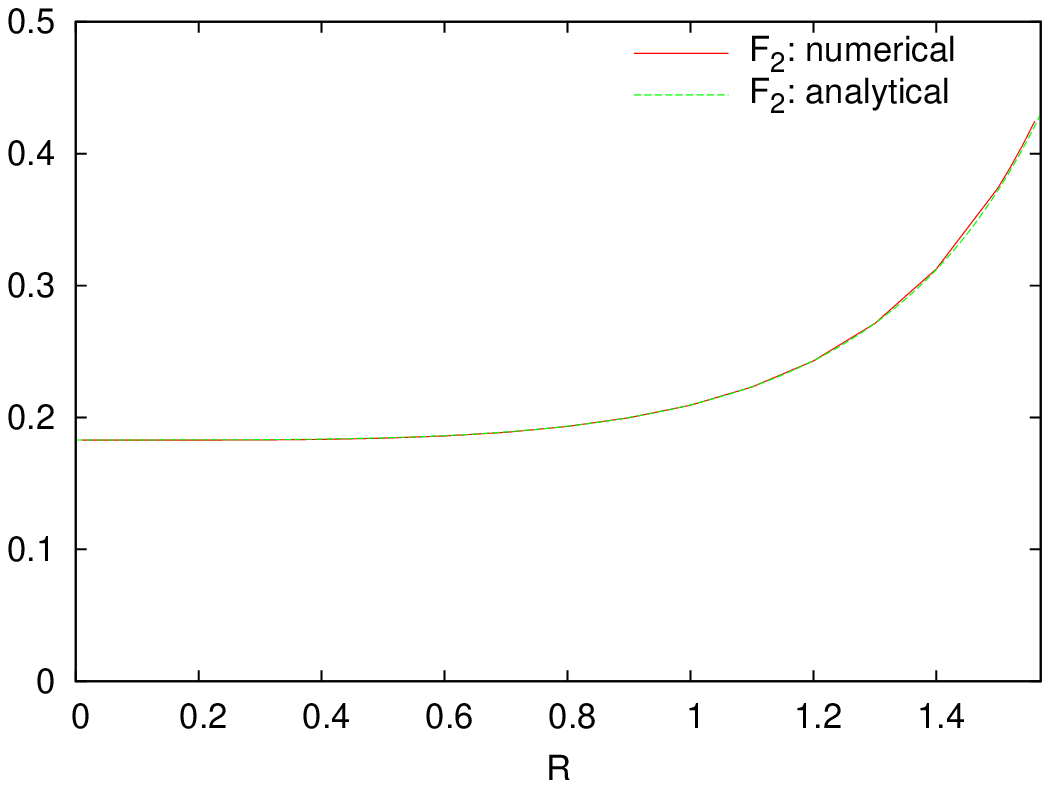,width=0.7\textwidth}
\caption{Full $R$-dependence of the two-gluon clustering
coefficient $\mc F_2$.\label{fig:F2Full}}}

Finally we note that the calculation of $\mc F_3^{(\ca)}$ and $\mc
F_4$ may be performed in the same way. Schematically we have:
\begin{equation}\label{eq:FnRInt}
\mc F_n (R) = \frac{1}{\pi^n} \int \prod_i^n d\eta_id\phi_i
\frac{1}{\cosh^2\eta_i-\cos^2\phi_i} \Xi_n(k_1,k_2,\cdots,k_n),
\end{equation}
where the step function $\Xi_n$ is expressed in terms of the
distances $d_{i}=\eta_i^2+\phi_i^2$ and $d_{ij} = \delta \eta_{ij}^2
+ \delta \phi_{ij}^2$, respectively replacing $\theta_i$ and
$\theta_{ij}$ in, e.g. eq. \eqref{eq:ClustFun3kt}. We provide full
numerical estimates for all coefficients in table
\ref{tab:ClusCoeffNumericalValues}.

\TABLE[ht]{
\centering
\begin{tabular}{|c|c|c|c|c|c|c|}
\hline
 $R$         & $0$      & $0.1$    & $0.4$    & $0.7$    & $1.0$    & $1.2$ \\
\hline\hline
 $\mc F_2$   & $0.183$  & $0.184$  & $0.184$  & $0.188$  & $0.208$  & $0.242$ \\
\hline
$\mc F_3$    & $-0.052$ & $-0.053$ & $-0.053$ & $-0.055$ & $-0.061$ & $-0.072$ \\
\hline
$\mc F_3^\ca$& $-0.028$ & $-0.029$ & $-0.029$ & $-0.029$ & $-0.030$ & $-0.031$ \\
\hline
$\mc F_4$    & $0.022$  & $0.023$  & $0.023$  & $0.023$  & $0.024$  & $0.027$ \\
\hline
\end{tabular}
\label{tab:ClusCoeffNumericalValues}
\caption{Estimates of the
clustering coefficients $\mc F_{n}$ for $n=2,3, 4$ at different
values of the jet radius $R$ in the $\kt$ and C/A algorithms. Recall
that $\mc F_2^\ca = \mc F_2^{\kt} (\equiv \mc F_2)$ and that $\mc
F_4^\ca$ has not been computed. Note that $\mc F_3^\ca$ and $\mc F_4$ are more difficult to evaluate numerically than $\mc F_3$ (and obviously $\mc F_2$). Thus their final answers are subjected to larger errors.}}

\end{document}